%% file: main.tex
\documentclass[runningheads,orivec]{llncs}

\newif\ifanonymous
\anonymousfalse

\usepackage{silence}
\usepackage[T1]{fontenc}
\usepackage[utf8]{inputenc}
\usepackage{graphicx}
\usepackage{adjustbox}
\usepackage{subcaption}
\usepackage{microtype}
\usepackage{enumitem}
\usepackage{csquotes}
\setlist[itemize,enumerate]{itemsep=0pt,parsep=0pt,topsep=0pt}
\usepackage{booktabs}

\usepackage{tikz}
\usetikzlibrary{arrows.meta, shapes.multipart, calc, backgrounds, fit}

\input{macros}
\ifanonymous\renewcommand{\nbc}[3]{}\fi

\usepackage{xspace}
\newcommand{\smartvote}{\emph{smartvote}\xspace}
\newcommand{\wahlomat}{\emph{Wahl-O-Mat}\xspace}

\usepackage{hyperref}

\usepackage[capitalise,noabbrev]{cleveref}

\begin{document}

\title{Comparing Dialogue Management Strategies with a Modular Platform for Conversational Agent Voting Advice Applications}

\titlerunning{Comparing Dialogue Management Strategies in CAVAAs}

\ifanonymous
\author{Anonymous Author(s)}
\authorrunning{Anonymous Author(s)}
\institute{Anonymous Institution}
% Used in the user study section (ethics approval).
\newcommand{\ethicscommittee}{the ethics committee of the authors' institution}
\else
\author{Selene Báez Santamaría \and
Patric Salvisberg \and
Fynn Bachmann}
\authorrunning{Báez Santamaría et al.}

\institute{University of Zurich, Zurich, Switzerland\\
\email{\{selene.baezsantamaria,patric.salvisberg,fynn.bachmann\}@uzh.ch}}
\newcommand{\ethicscommittee}{the ethics committee of the Faculty of Business, Economics and Informatics of the University of Zurich under grant ID 2026-069}
\fi

\maketitle

\begin{abstract}
\input{content/abstract}
\end{abstract}

\input{content/introduction}
\input{content/related_work}

\input{content/chatbot}
\input{content/experimental_setup}
\input{content/results}
\input{content/discussion}
\input{content/limitations}
\input{content/conclusion}

\begin{credits}
\subsubsection{\ackname}
We acknowledge the Swiss National Science Foundation (SNSF), which funds this research with grant ID CRSII5-205975.
Furthermore, we are grateful to the team from \textit{Politools} for providing the \smartvote data.
A \textit{Norstat Academic grant} supported the recruitment of participants in the user study.
We also thank Thilo I. Dieing and Abraham Bernstein for advice throughout the research process, Naomi Kamoen for helpful feedback on an early draft of this paper, and the anonymous reviewers for their valuable comments.

\subsubsection{\discintname}
The authors have no competing interests to declare.

\subsubsection{Data Availability.}
The data and analysis code of this paper, together with the supplementary material, are available at \href{https://github.com/fsvbach/conversations-cavaa-paper}{\nolinkurl{fsvbach/conversations-cavaa-paper}}.
The code for the platform deployment is available at \href{https://github.com/fsvbach/aqvaa-platform}{\nolinkurl{fsvbach/aqvaa-platform}} and \href{https://github.com/fsvbach/aqvaa-chatbot}{\nolinkurl{fsvbach/aqvaa-chatbot}}.
\end{credits}

% ---- Bibliography ----
\bibliographystyle{splncs04}
\bibliography{bib/references}

\end{document}

%% file: macros.tex
\usepackage{xcolor}
\usepackage{amssymb}

\definecolor{selcolor}{rgb}{0.53,0.30,1.00}
\definecolor{fynncolor}{rgb}{0.5,0,0}
\definecolor{patriccolor}{rgb}{0,0.5,0}
\definecolor{todocolor}{rgb}{1,0,0}
\definecolor{placeholdercolor}{rgb}{1,.5,.2}

\newcommand{\nbc}[3]{
		{\colorbox{#3}{\bfseries\sffamily\scriptsize\textcolor{white}{#1}}}
		{\textcolor{#3}{\sf\small$\blacktriangleright$\textit{#2}$\blacktriangleleft$}}
}

\definecolor{addedcolor}{rgb}{0.0,0.45,0.65}

\newcommand{\model}[1]{\textsf{\small#1}}

\newcommand{\ci}[3]{$g=#1$ $[#2,\,#3]$}
\renewcommand{\ci}[3]{$g=#1$}

\makeatletter
\AtBeginDocument{% llncs defines \doi at begin-document, so wrap it here.
  \let\cavaa@doi\doi
  \def\doi#1{\cavaa@doi{#1}\@ifnextchar,{\cavaa@doi@comma}{}}}
\def\cavaa@doi@comma,{\@ifnextchar\url{\cavaa@gobble@url}{, }}
\def\cavaa@gobble@url{\begingroup
  \let\do\@makeother\dospecials \catcode`\{=1 \catcode`\}=2
  \cavaa@gobble@url@i}
\def\cavaa@gobble@url@i\url#1{\endgroup}
\makeatother
\usepackage{etoolbox}  % \ifblank, \ifstrequal
\definecolor{chatbotfill}{RGB}{235,238,243}
\definecolor{chatuserfill}{RGB}{221,236,226}
\definecolor{chattag}{RGB}{158,42,42}
\newcounter{chatturn}
\newlength{\chatgap}
\newlength{\chattagextra}  % extra gap above a bubble that carries a tag
\makeatletter
\newlength{\chat@shift}
\newenvironment{chat}{%
  \setcounter{chatturn}{0}%
  \begin{tikzpicture}[inner sep=0pt, outer sep=0pt, baseline=(chat0L),
      bubble/.style={rounded corners=4pt, inner sep=3pt, align=left,
                     font=\scriptsize\sffamily, text width=0.9\linewidth-6pt},
      bot/.style={bubble, fill=chatbotfill},
      user/.style={bubble, fill=chatuserfill},
      tag/.style={font=\scriptsize\sffamily\bfseries, text=white, fill=chattag,
                  rounded corners=2pt, inner xsep=2.5pt, inner ysep=0.7pt,
                  text height=1.35ex, text depth=0.35ex}]
    \coordinate (chat0L) at (0,0);
    \coordinate (chat0R) at (\linewidth,0);
    \path (chat0L) -- (chat0R);
}{\end{tikzpicture}}
\newcommand{\chat@bubble}[5]{%
  \edef\chat@prev{chat\thechatturn#3}%
  \stepcounter{chatturn}%
  \ifblank{#4}{\setlength{\chat@shift}{\chatgap}}{\setlength{\chat@shift}{\dimexpr\chatgap+\chattagextra\relax}}%
  \node[#1, anchor=#2] (chat\thechatturn) at ([yshift=-\chat@shift]\chat@prev) {#5};
  \coordinate (chat\thechatturn L) at (chat\thechatturn.south -| chat0L);
  \coordinate (chat\thechatturn R) at (chat\thechatturn.south -| chat0R);
  \ifblank{#4}{}{%
    \ifstrequal{#3}{L}%
      {\node[tag, anchor=south east] at ([xshift=-4pt, yshift=-2pt]chat\thechatturn.north east) {#4};}%
      {\node[tag, anchor=south west] at ([xshift=4pt, yshift=-2pt]chat\thechatturn.north west) {#4};}}%
}
\newcommand{\botsays}[2][]{\chat@bubble{bot}{north west}{L}{#1}{#2}}
\newcommand{\usersays}[2][]{\chat@bubble{user}{north east}{R}{#1}{#2}}
\makeatother

%% file: content/abstract.tex
Voting Advice Applications (VAAs) rely on questionnaires to recommend parties or candidates to voters.
However, voters perceive some of the VAA statements as too complex to answer meaningfully.
In this paper, we present a platform that integrates a conversational agent into a VAA to support users in asking questions, reflecting on their preferences, and giving better-informed responses.
The LLM-based chatbot contains specialized modules for guardrails, query routing, factual retrieval, and per-turn stance classification. It supports two dialogue management strategies: a reactive assistant and a proactive interviewer.
We evaluate the modules on dedicated datasets and compare the two strategies in a user study with 75 participants.
Users of the assistant mainly ask factual questions, whereas users of the interviewer mainly answer the chatbot's yes/no questions and state their own opinion more often.
At the same time, participants tend to rate the interviewer as revealing a political stance of its own and as steering them toward an answer more than the assistant, with no conclusive differences in usability ratings, political interest, or vote certainty.
The open-source platform is intended to support the community in investigating how different dialogue management strategies in conversational VAAs affect users' political knowledge and engagement.

\keywords{Voting Advice Applications \and Conversational Agents \and Stance Classification \and Dialogue Management \and Retrieval-Augmented Generation}

%% file: content/introduction.tex
\section{Introduction}

In multi-party democracies, Voting Advice Applications (VAAs) contribute to political education by recommending candidates or parties to voters through well-designed surveys~\cite{garzia_voting_2019,tromborg_candidates_2023}.
% \sel{R3: Scandinavia not a country}
Recently, these tools have gained considerable relevance: the German \wahlomat was used over 26 million times in 2025, \smartvote in Switzerland attracted 20\% of the electorate in the 2023 election, and in Finland and Norway, VAAs reach up to 68\% of the voting-age population~\cite{garzia_how_2026}.

% \sel{R3: satisfying}
Despite this popularity, traditional VAAs suffer from semantic and pragmatic comprehension problems that lead users to satisficing behavior, giving answers that are good enough rather than carefully considered~\cite{kamoen_dont_2017,baka_neither_2012}.
Users often lack enough background on specific policies to answer the statements meaningfully.
Recent work addresses this with Conversational Agent Voting Advice Applications (CAVAAs), where users can ask clarification questions or reflect on their intuitions before giving a better-informed response~\cite{kamoen_need_2022,dieing_aidriven_2024,zhu_learn_2025}.
In most CAVAAs, users answer the same VAA questionnaire through the explicit vote buttons while the chatbot adds conversational support.
% \sel{introduce gap}
Whether the chatbot should instead lead the conversation, as in AI conversational interviewing~\cite{wuttke_ai_2025}, has received little attention in CAVAAs based on Large Language Models (LLMs).

% \sel{R3: motivation/purpose statement}
In this work, we investigate how a modular, LLM-based conversational agent can support political education in a VAA, and how the dialogue management strategy affects how users interact with and perceive it.
% \sel{R3: map contributions to research questions}
We address two research questions.
\textbf{RQ1:} How well do LLM-based modules for guardrails, query routing, retrieval, and stance classification perform in a CAVAA, on benchmarks and in live conversations?
\textbf{RQ2:} How do users interact with and perceive a CAVAA under a reactive and a proactive dialogue management strategy?

\begin{figure}[t]
    \centering
    \input{content/pipeline}
    \caption{System architecture with processing steps and data flow. 
    Input data (blue), knowledge base (orange), and chatbot settings (teal) feed into processing modules, producing the output shown to the user (purple).
    Solid arrows denote control flow, dashed arrows data flow, and the dotted red arrow the guardrail rejection.}
    \label{fig:architecture}
\end{figure}
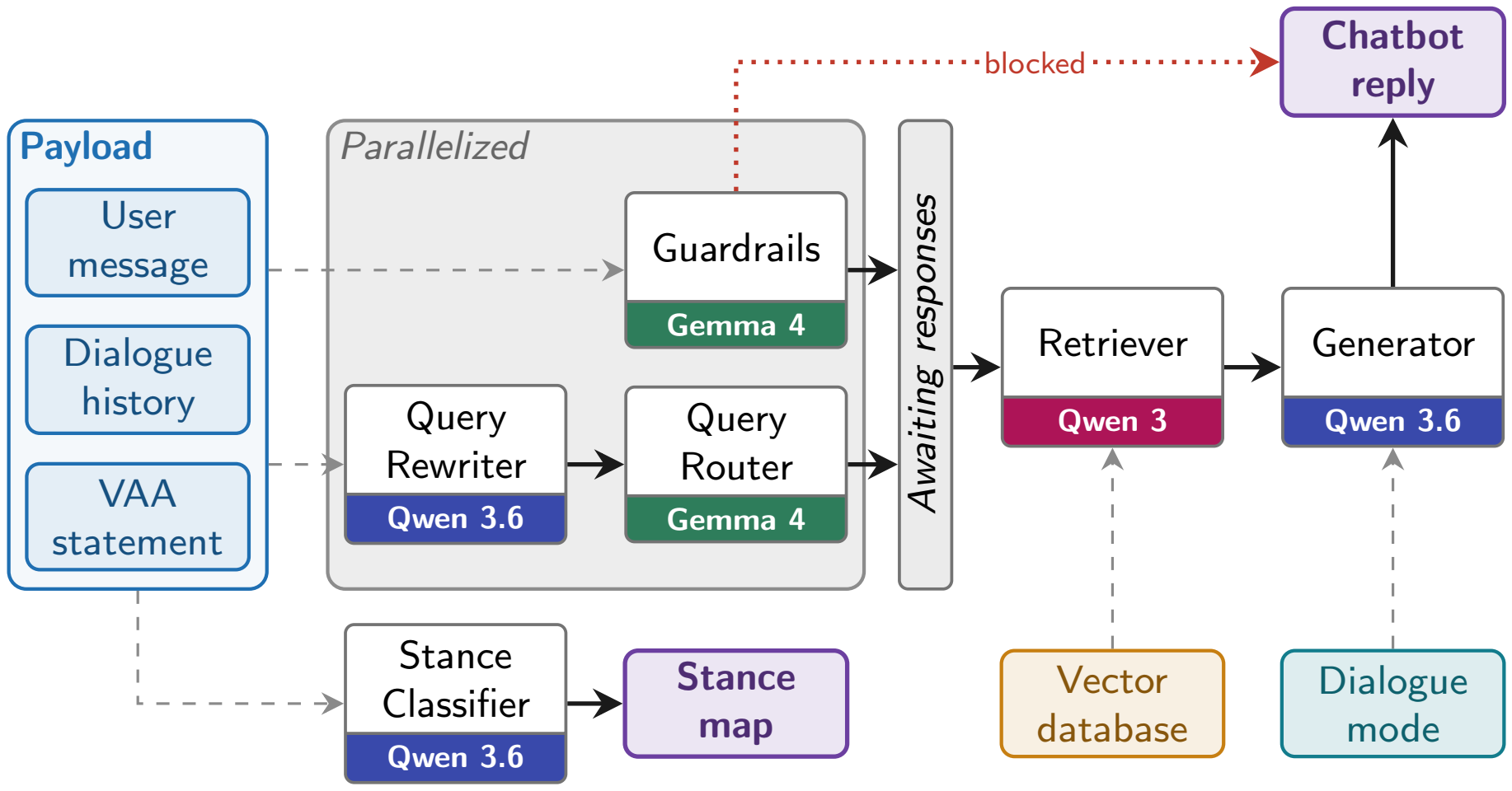

To address RQ1, we build on recent CAVAAs that combine open-weight LLMs with Retrieval-Augmented Generation (RAG)~\cite{dieing_traditional_2025,schiele_voting_2024}, design a modular pipeline of guardrails, query routing, factual retrieval, and stance classification (\cref{fig:architecture}), and evaluate its modules on dedicated datasets.
On top of this pipeline, the platform supports two dialogue management strategies: a reactive assistant that answers questions next to the traditional VAA interface, and a proactive interviewer that integrates the VAA into the chat and elicits user opinions through follow-up questions.

To address RQ2, we compare the two strategies in a user study with 75 participants in the context of Swiss politics.
We find that users of the assistant mainly ask factual questions, whereas users of the interviewer mainly answer the chatbot's yes/no questions and state their own opinion more often.
At the same time, participants tend to rate the interviewer as revealing a political stance of its own and as steering them toward an answer more often than the assistant.

% Our contributions are two-fold: for RQ1, an open-source CAVAA platform based on open-weight LLMs with RAG, per-turn stance classification, safety guardrails, and evaluation datasets for its modules; for RQ2, exploratory results that compare a reactive assistant and a proactive interviewer in terms of conversation patterns and user perception.
Together, these exploratory results and the open-source platform provide a basis for further research on CAVAA design and on conversational agents for political education.

%% file: content/pipeline.tex
% CAVAA system pipeline -- reusable TikZ body.
%
% Designed to fit the LNCS text width (12.2cm) at natural size: no
% \resizebox needed. Vertical space is used freely instead.
%
% The host document's PREAMBLE must load tikz and these libraries (loading them
% mid-document inflates the bounding box, so they are NOT loaded here):
%   \usepackage{graphicx}   % \rotatebox for the join bar
%   \usepackage{tikz}
%   \usetikzlibrary{arrows.meta, shapes.multipart, calc, backgrounds, fit}
%
% Arrow semantics (explain in the caption): solid = control flow,
% dashed grey = data, dotted red = guardrail rejection.
%
\begin{tikzpicture}[
  x=1cm, y=1cm, font=\sffamily,
  % --- process box: fixed-size title block + coloured model strip ---
  proc/.style={
    rectangle split, rectangle split parts=2, rectangle split part align=center,
    draw=black!55, line width=0.7pt, rounded corners=2pt,
    rectangle split part fill={white, cQwen36},
    inner sep=2.4pt, font=\sffamily\small},
  gem/.style={proc, rectangle split part fill={white, cGemma}},
  qwn36/.style={proc, rectangle split part fill={white, cQwen36}},
  qwn3/.style={proc, rectangle split part fill={white, cQwen3}},
  % --- join bar: waits for guardrails + router before retrieval ---
  joinbar/.style={rectangle, rounded corners=2pt, draw=black!55, line width=0.7pt,
                  fill=black!8, minimum width=4.2mm, inner sep=0pt,
                  font=\sffamily\footnotesize\itshape},
  % --- inputs / settings / outputs: plain rounded rectangles ---
  box/.style={rounded corners=3pt, line width=0.8pt, text width=16mm, align=center,
              inner sep=2.6pt, minimum height=8.5mm, font=\sffamily\footnotesize},
  inp/.style={box, draw=cInput, fill=cInput!12, text=cInput!72!black},
  store/.style={box, draw=cStore, fill=cStore!12, text=cStore!68!black},
  mode/.style={box, draw=cMode, fill=cMode!14, text=cMode!78!black},
  outp/.style={box, draw=cOut, line width=1pt, fill=cOut!13, text=cOut!72!black,
               font=\sffamily\footnotesize\bfseries},
  % --- groups (payload container, parallel region) ---
  grp/.style={rectangle, rounded corners=4pt, inner sep=1.3mm, line width=0.8pt},
  grplabel/.style={anchor=north west, inner sep=2.5pt, font=\sffamily\footnotesize},
  % --- arrows (blocked matches flow in size, stays dotted/red) ---
  flow/.style={-{Stealth[length=2.4mm,width=2.4mm]}, line width=1pt, draw=cFlow},
  data/.style={-{Stealth[length=1.8mm,width=1.8mm]}, line width=0.6pt, draw=cData, dashed},
  block/.style={-{Stealth[length=2.4mm,width=2.4mm]}, line width=1pt, draw=cBlock, dotted},
]

% --- model labels (override via \providecommand before \input, or edit here) -
\providecommand{\cvstrip}[1]{{\sffamily\scriptsize\bfseries\textcolor{white}{#1}}}
\providecommand{\cvtitle}[1]{\parbox[c][6.8mm][c]{16mm}{\centering #1}}
\providecommand{\mGuard}{Gemma 4}
\providecommand{\mRoute}{Gemma 4}
\providecommand{\mRewrite}{Qwen 3.6}
\providecommand{\mRetrieve}{Qwen 3}
\providecommand{\mGenerate}{Qwen 3.6}
\providecommand{\mStance}{Qwen 3.6}
% --- palette ---
\definecolor{cInput}{HTML}{1F6FB2}   % inputs (blue)
\definecolor{cStore}{HTML}{C77F12}   % knowledge base (orange)
\definecolor{cMode}{HTML}{127C8C}    % mode (teal -- distinct from store)
\definecolor{cOut}{HTML}{6C3FA0}     % outputs (violet)
\definecolor{cGemma}{HTML}{2E7D5B}   % model family A (green)
\definecolor{cQwen36}{HTML}{3949AB}  % model family B (indigo)
\definecolor{cQwen3}{HTML}{AD1457}   % model family C (magenta)
\definecolor{cFlow}{HTML}{1A1A1A}    % control flow
\definecolor{cData}{HTML}{8A8A8A}    % data / context flow
\definecolor{cBlock}{HTML}{C0392B}   % blocked path

% ===== grid ==========================================================
% columns: payload | rewriter | guardrails+router | join | retriever | generator
\def\xP{0}    \def\xA{2.55} \def\xB{4.8} \def\xJ{6.32} \def\xR{7.82} \def\xG{10.07}
% rows: reply lane | parallel row A | main chain | parallel row B | bottom row
\def\yTop{2.45} \def\yA{0.78} \def\yB{-0.78} \def\yS{-2.7}
% shared vertical extent of the payload box, parallel box and join bar
\def\yBoxTop{1.85} \def\yBoxBot{-1.65}

% ===== payload (all three inputs, one container) ====================
\node[inp] (userq)   at (\xP, 1.0) {User message};
\node[inp] (history) at (\xP,-0.1) {Dialogue history};
\node[inp] (vaaq)    at (\xP,-1.2) {VAA statement};

% ===== parallel region: guardrails || rewriter -> router ============
\node[gem]   (guard)   at (\xB,\yA) {\cvtitle{Guardrails}     \nodepart{two} {\cvstrip{\mGuard}}};
\node[qwn36] (rewrite) at (\xA,\yB) {\cvtitle{Query Rewriter} \nodepart{two} {\cvstrip{\mRewrite}}};
\node[gem]   (route)   at (\xB,\yB) {\cvtitle{Query Router}   \nodepart{two} {\cvstrip{\mRoute}}};

% ===== join -> retrieval -> generation ==============================
\node[joinbar, fit={(\xJ,\yBoxTop) (\xJ,\yBoxBot)}, inner ysep=1.3mm] (join) {};
\node[font=\sffamily\footnotesize\itshape, inner sep=0pt] at (join) {\rotatebox{90}{Awaiting responses}};
\node[qwn3]  (retrieve) at (\xR,0) {\cvtitle{Retriever} \nodepart{two} {\cvstrip{\mRetrieve}}};
\node[qwn36] (generate) at (\xG,0) {\cvtitle{Generator} \nodepart{two} {\cvstrip{\mGenerate}}};

% ===== bottom row: stance branch + knowledge base + settings ========
\node[qwn36] (stance) at (\xA,\yS) {\cvtitle{Stance Classifier} \nodepart{two} {\cvstrip{\mStance}}};
\node[outp]  (smap)   at (\xB,\yS) {Stance map};
\node[store] (vdb)    at (\xR,\yS) {Vector database};
\node[mode]  (mode)   at (\xG,\yS) {Dialogue mode};

% ===== output ======================================================
\node[outp]  (reply) at (\xG,\yTop) {Chatbot reply};

% ===== containers (background so nodes stay on top) =================
\begin{scope}[on background layer]
  \node[grp, draw=cInput, fill=cInput!5,
        fit={(userq.west |- 0,\yBoxTop) (userq.east |- 0,\yBoxBot)}] (payload) {};
  \node[grplabel, text=cInput] at (payload.north west) {\bfseries Payload};
  \node[grp, draw=black!45, fill=black!7,
        fit={(rewrite.west |- 0,\yBoxTop) (route.east |- 0,\yBoxBot)}] (par) {};
  \node[grplabel, text=black!60] at (par.north west) {\itshape Parallelized};
\end{scope}

% ===== connectors ===================================================
% ----- control flow -----
\draw[flow] (rewrite)    -- (route);
\draw[flow] (guard.east) -- (join.west |- guard.east);
\draw[flow] (route.east) -- (join.west |- route.east);
\draw[flow] (join.east |- retrieve.west) -- (retrieve.west);
\draw[flow] (retrieve)   -- (generate);
\draw[flow] (generate.north) -- (reply.south);
\draw[flow] (stance)     -- (smap);

\begin{scope}[on background layer]
% ----- data inputs (dashed) -----
\draw[data] (payload.east |- guard.west)   -- (guard.west);
\draw[data] (payload.east |- rewrite.west) -- (rewrite.west);
\draw[data] (payload.south) |- (stance.west);
\draw[data] (vdb.north)  -- (retrieve.south);
\draw[data] (mode.north) -- (generate.south);
\end{scope}

% ----- guardrail rejection / short-circuit -> reply -----
\draw[block] (guard.north) |- (reply.west);
\node[font=\sffamily\scriptsize, text=cBlock, fill=white, inner sep=1pt]
  at (7.2,\yTop) {blocked};

\end{tikzpicture}

%% file: content/related_work.tex
\section{Related Work}

% Exisitng and growing CAVAA literature
CAVAAs were introduced in~\cite{kamoen_need_2022}, where a chatbot embedded in a VAA increased perceived and factual political knowledge compared to a traditional interface.
Subsequent studies confirmed these benefits~\cite{vanzanten_voting_2024,velez_chatbot_2025,zhu_learn_2025}.
To further improve these effects, a growing literature now focuses on CAVAA architectures~\cite{vassos_now_2024}, design~\cite{hankel_hi_2025}, and safety~\cite{dieing_traditional_2025}.

% RAG? 
Investigating user engagement in CAVAAs, a comparison of chatbot designs found that structured and semi-structured conversational formats were more engaging than unstructured ones~\cite{kamoen_conversational_2022}.
However, the rule-based dialogue management in these studies limited their scope and scalability.
LLMs can overcome these constraints~\cite{dieing_aidriven_2024}, but are prone to hallucination and were found to sometimes deviate from the parties' stated positions~\cite{dormuth_cautionary_2026}.
More recent systems therefore ground responses in party manifestos through RAG~\cite{lewis_retrievalaugmented_2020,shuster_retrieval_2021,vassos_now_2024,schiele_voting_2024}.

% Proactive vs Reactive
In parallel, recent work proposes AI conversational interviewing to scale semi-structured interviews with LLMs~\cite{wuttke_ai_2025,wuttke_ai_2026}.
Here, the model asks open-ended questions on political topics and probes the answers with follow-up questions while adhering to a predefined interview guide.
Such interviewers can also code the open-ended answers during the conversation~\cite{barari_aiassisted_2026}, for example detecting stances~\cite{kucuk_stance_2021,walker_promises_2025} that map user messages back onto the VAA statement.
This demonstrates the potential of proactive dialogue for eliciting opinions in the respondents' own words and offers an opportunity to further personalize the CAVAA experience.
Such a framework could lay the foundation for conversation-based political education, where users inform themselves and convey their position within the same conversation.
However, it was found that a proactive chatbot, which asks follow-up questions, correlated with earlier dropout of users~\cite{vanderweijden_typology_2026}.
Our study provides first evidence of how proactive and reactive chatbots are perceived in LLM-driven CAVAAs, and which conversational patterns drive this perception.

%% file: content/chatbot.tex
\section{Platform}
\label{sec:chatbot}

To evaluate different CAVAA designs, we develop a modular VAA platform that integrates a conversational agent.
During users' conversation with the chatbot, each user message is processed through a pipeline of LLM-based modules: guardrails, query routing, factual retrieval, and stance classification (\cref{fig:architecture}).
We benchmark several LLM families and sizes on each module's own evaluation dataset, including
Qwen's \model{Qwen2.5-3B}, \model{Qwen3-4B}, \model{Qwen3-8B}, \model{Qwen3.5-2B}, \model{Qwen3.5-4B}, and \model{Qwen3.6-35B-A3B-FP8};
Google's \model{gemma-4-E2B-it} (5B model) and \model{gemma-4-E4B-it} (8B model);
and Mistral's \model{Ministral-3-14B-Instruct-2512}.

Based on the evaluation results, the final configuration uses the following open-weight models:
\model{gemma-4-E4B-it} is used for guardrails and query routing.
\model{Qwen3.6-35B-A3B-FP8} handles query rewriting, response generation, and stance classification.
\model{Qwen3-Embedding-8B} performs dense retrieval.
This mix balances task performance and speed, ensuring near-instant responses and a satisfactory user experience.
Below, we describe each module and the corresponding evaluation in more detail.

\subsection{Guardrails}
\label{sec:guardrails}
% What the module does
Guardrails perform four separate checks, run concurrently with the same model, namely whether the message is off-topic for Swiss politics (Prompt~1)\footnote{All numbered prompts are in the supplementary material, see Data Availability.}, contains insults or asks for the chatbot's opinion on a controversial topic (Prompt~2), explicitly asks for a vote recommendation (Prompt~3), or otherwise asks for the chatbot's own opinion (Prompt~4).
A message is blocked if any of the four checks fires.

% Dataset
We build a shared evaluation dataset by hand-crafting 118 German test messages across 16 categories, twelve adapted from~\cite{dieing_aidriven_2024} and four tailored to our use case (candidate position, candidate vs party, cross-party overview, party comparison).
We augment the dataset by paraphrasing each message ten times using \model{Qwen3-8B}, covering lexical, syntactic, informal, formal, and implicit reformulations, which yields 1,298 messages.
Four of the 16 categories correspond to the messages that the guardrails should block (26 seed messages and 286 with paraphrases).

% Performance
The guardrails block 92\% of the messages that should be blocked (recall), and 91\% of the blocked messages should indeed be blocked (precision), for an F1 score of 0.91.
They wrongly block 2.7\% of the legitimate messages (27 of 1,012), mostly questions that ask for a party's or candidate's position in the words of an opinion request, such as \enquote{What does <candidate> think about it?}.

\subsection{Query Routing}
\label{sec:routing}
% What the module does
Query routing classifies the user message to determine which sources should be used to answer (Prompt~5).
The router returns a set of sources with optional party or candidate filters, or an empty set for messages that need no retrieval.
Before routing, we condense the dialogue history and the last user message into a single self-contained query (Prompts~6 and~7).

% Dataset
We evaluate query routing on the same shared dataset.
Twelve categories contain user messages that should be routed (92 seed messages and 1,012 with paraphrases), while the rest require no retrieval.
Each category is associated with predefined sources and filters.

% Performance
A routing decision counts as correct only if the predicted set of sources and filters matches the expected set exactly.
89\% of the messages are routed correctly, reaching a macro-average F1 score of 0.94 over the source sets.
For the messages that need no retrieval, an empty set is correctly returned in 97\% of the cases.

\subsection{Factual Retrieval}
\label{sec:retrieval}
% What the module does
We use a knowledge base built from official party documents (e.g. manifestos, position papers, and statements published on the parties' websites) together with candidate data from \smartvote~\cite{burst_manifestoberta_2024,politools_daten_2023}.
The final knowledge base covers eight parties across 85 documents.
We convert each document to Markdown and split it along its heading hierarchy, keeping passages of up to 512 characters.
This yields 6,381 passages.

% Dataset
To evaluate retrieval, we use the five routing categories that require retrieval from the party documents: party position, party comparison, party rationale, cross-party overview, and party ideology.
These categories contain 62 seed messages and 682 with paraphrases.
For each seed message, we manually label the 20 passages ranked highest by \model{Qwen3-Embedding-8B} as relevant or irrelevant (1,205 labels, 276 relevant)\footnote{The evaluation was run on an earlier version of the knowledge base with ten documents from seven parties, split into 1,550 passages.}.
Paraphrases inherit the labels of their seed.

% Performance
Retrieval finds on average 37\%, 54\%, and 69\% of a message's relevant passages within the top 5, 10, and 20 results (recall@$k$).
For 93\% of the messages, at least one relevant passage is among the top 20, and the first relevant passage has a median rank of two.
The retrieved passages are verbalized into a response (Prompts~8~to~13). 
% We evaluate response generation with \model{claude-haiku-4-5} as an LLM judge on 82 cases, reaching a mean faithfulness of 0.99, mean relevancy of 0.97, and a 96\% pass rate (both scores above 0.7).

\subsection{Stance Classification}
\label{sec:stance}
% What the module does
We incorporate stance classification at each conversational turn (Prompt~14).
We classify whether the user has expressed a stance on the current VAA statement, and if so, in which direction.
This allows us to track changes in the user's position as the conversation progresses.

% Dataset
To evaluate stance classification, we use data from the Swiss VAA \smartvote~\cite{politools_daten_2023}.
The data include responses from 1,029 candidates to 60 political statements.
Each candidate rated each statement on an ordinal scale with four categories (disagree, rather disagree, rather agree, agree).
In 16\% of the candidate-statement pairs, candidates justified their answer in a free-text comment.
We use these 9,627 candidate comments as input and the answers they provided as ground truth.
In addition to the four response categories, we introduce an \enquote{irrelevant} category.
This is needed as not all user messages during a conversation directly express a stance toward the VAA statement, while the \smartvote candidate comments justify their stance and are therefore naturally on-topic.

% Performance
The stance classification module achieves 56\% accuracy on exact matches across the four categories; comments classified as irrelevant count as errors.
A binary accuracy collapses the scale into agree and disagree.
Under this metric, the module achieves an accuracy of 80\%.
For comparison, a latent space model based on all candidates' answers reaches around 82\% imputation accuracy on this dataset~\cite{bachmann_fast_2024}.

%% file: content/experimental_setup.tex
\section{User Study}
\label{sec:exp_setup}

To evaluate the platform from a user perspective, we conducted an online study comparing two dialogue management strategies: a reactive assistant and a proactive interviewer.
Both strategies run on the same pipeline and differ in their system prompts (Prompts~8~to~13) and in how the chatbot is integrated into the VAA interface.
Below, we describe the two conditions, the experimental setup, and how we annotate the resulting conversations.

\subsection{Dialogue Management Strategies}
\label{sec:conditions}

\begin{figure}[!t]
    \centering
    % raw Chrome captures (390 x 1060 px, DPR 3, root font size 22 px); the trim cuts the "EN | DE" header and the page padding below the card; at 0.48\textwidth per panel the figure is too tall for the top of a page and adds a 19th page
    \begin{subfigure}[t]{0.42\textwidth}
        \centering
        \adjincludegraphics[width=\linewidth, trim={0 {.0236\height} 0 {.0623\height}}, clip]{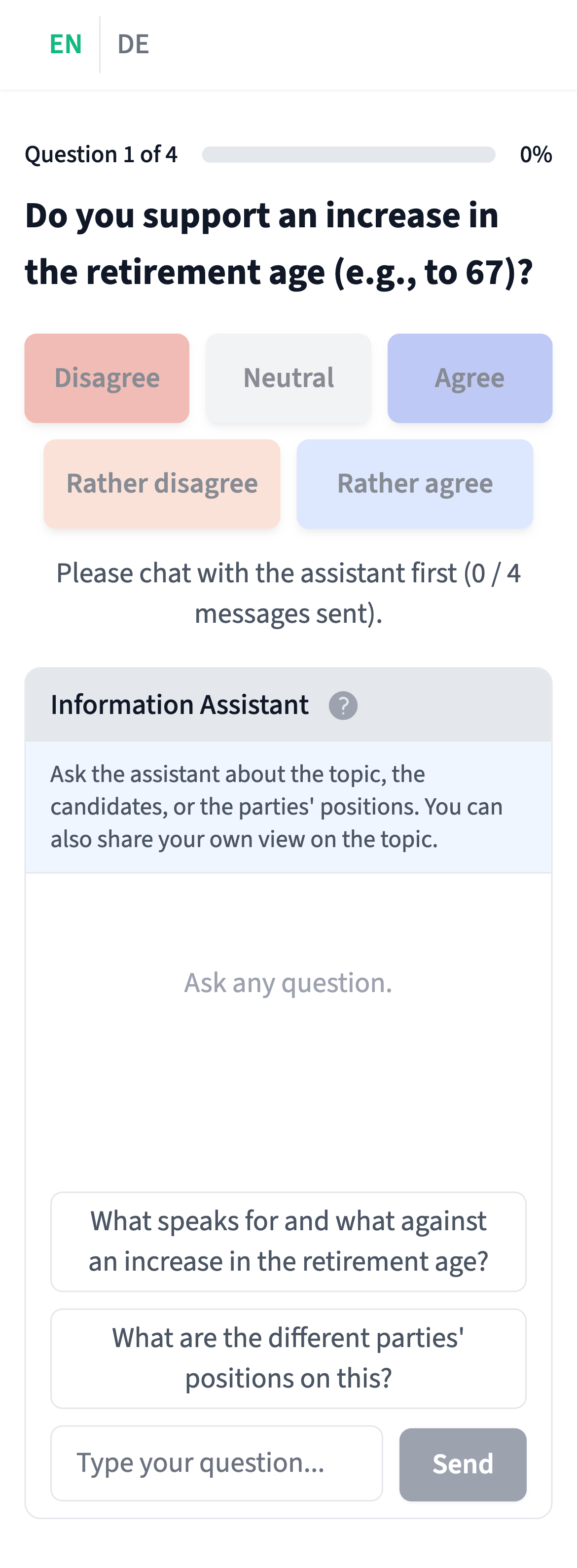}
        \caption{Assistant condition, with the VAA statement, answer buttons, and the chat interface with suggested messages.}
        \label{fig:assistant}
    \end{subfigure}
    \hspace{0.04\textwidth}
    \begin{subfigure}[t]{0.42\textwidth}
        \centering
        \adjincludegraphics[width=\linewidth, trim={0 {.0236\height} 0 {.0623\height}}, clip]{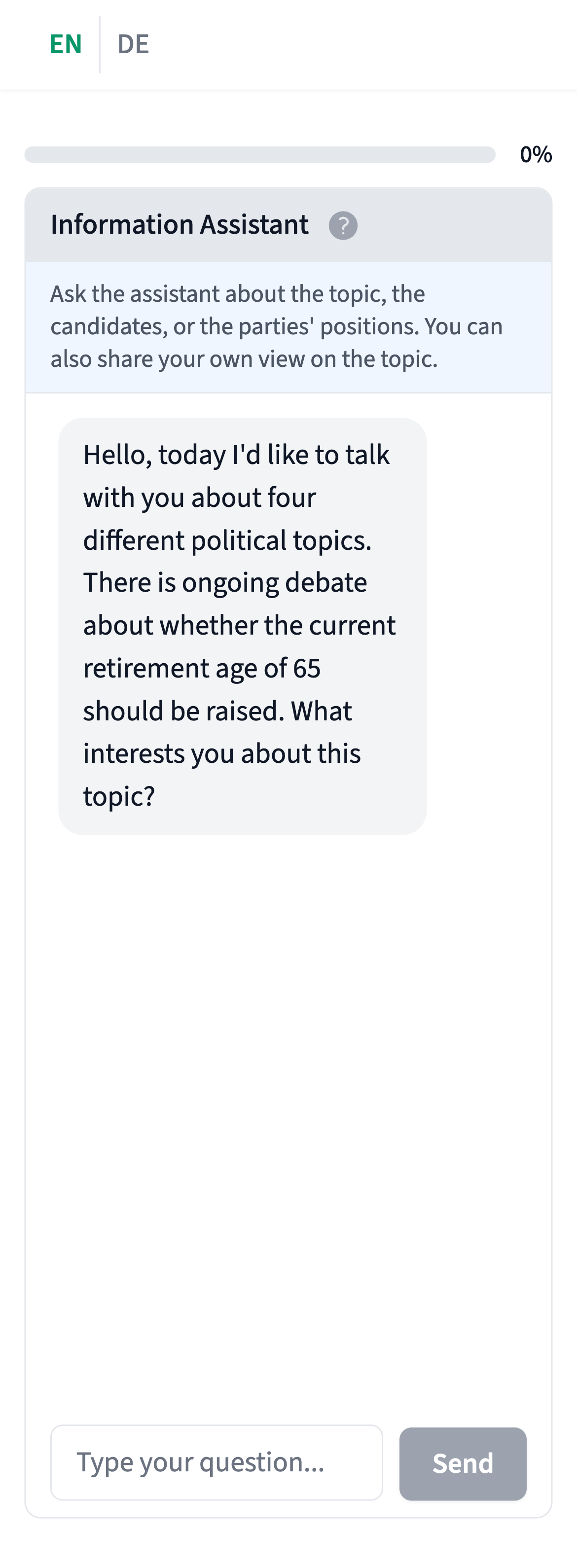}
        \caption{Interviewer condition, with the opening message and chat interface.}
        \label{fig:interview}
    \end{subfigure}
    \caption{Screenshots of the mobile interface.}
    \label{fig:ui}
\end{figure}

% Assistant 
In the assistant condition, users see the traditional VAA interface, consisting of the VAA statement and answer buttons, extended by a chat interface (\cref{fig:assistant}).
The chatbot responds to messages related to the VAA statement at hand, providing factual information on demand~\cite{kamoen_need_2022}.
Following evidence that semi-structured interfaces lead to better engagement than open-ended text input alone~\cite{kamoen_conversational_2022}, we offer predefined question suggestions alongside the free-text field.
The two predefined questions are hand-crafted for the first VAA statement.
In addition to providing convenient starting points, they are designed to help users form an accurate mental model of the chatbot's capabilities by showing the types of questions that can be asked.

% Interviewer
In the interviewer condition, the VAA is fully integrated into the chat interface, and the chatbot proactively guides users to reflect on and articulate their political preferences~\cite{zhu_learn_2025,vanderweijden_typology_2026}.
An introductory message explains the system and opens a discussion of the topic underlying the current VAA statement (\cref{fig:interview}).
Each chatbot response ends with a follow-up question that further elicits the user's stance on the topic.
Once the topic has been sufficiently discussed, the user is asked to confirm the inferred stance through answer buttons rendered within the chat.

\subsection{Experimental Setup}
\label{sec:setup}
% Recruitment
We recruited participants through the market research company Norstat and randomly assigned them to one of the two conditions.
All participants were eligible to vote in a Swiss canton.
The study ran over two days in July 2026 and was approved by \ethicscommittee.

% Flow
The study consisted of a pre-survey, the CAVAA platform, and a post-survey.
In total, 75 participants completed the pre-survey and opened the platform.
% Pre-survey
The pre-survey collected demographics, political interest, vote certainty, political efficacy, attitude toward VAAs, and AI literacy~\cite{grassini_psychometric_2024}.
% Topics

On the platform, participants were shown four \smartvote statements in German, covering retirement age, educational equality, nuclear power, and relations with the EU.
% Stop condition
In both conditions, participants had to exchange at least four messages not blocked by the guardrails before they could rate their agreement with each statement.
% VAA part
After the fourth statement, participants received candidate recommendations together with their position on a two-dimensional political map, computed from their answers.

% Post-survey
The post-survey measured perceived factuality, perceived political stance of the chatbot, perceived manipulation~\cite{dillard_nature_2005}, and technology acceptance (TAM)~\cite{davis_perceived_1989}.
It repeated political interest, vote certainty, and efficacy from the pre-survey and closed with two open questions about the user's experience with the chatbot.

\subsection{Conversation Annotation}
\label{sec:annotation}
% Discarded
We manually inspect all conversations and discard those in which the participant did not engage with the task, for example by sending only nonsensical messages.
% Manual annotation: user turns
We annotate every user turn with its stance toward the VAA statement on a five-point scale.
Turns without a stance receive one of the following five dialogue acts: general statement, question about content, question about the chatbot, command, or answer to the chatbot.
We regard any user turn annotated with a stance as expressing a \enquote{user opinion}.

% LLM as judge: chatbot turns
Chatbot turns are tagged by \model{Qwen3.6-35B-A3B-FP8} in an LLM-as-a-judge setup (Prompt~16).
The tags cover five dialogue acts (yes/no question, open question, opinion, statement, or answer), context breaks (failing to build on prior turns), stance leaks (expressing a political position of its own), and directive framing (steering the user toward a conclusion or emphasizing one side).
% LLM as judge: user turns
For user turns manually annotated as a question, the judge additionally classifies the question's form as yes/no or open-ended.
% Dialogue sequences
Combining the user and chatbot tags, we distinguish prompted from unprompted opinions (given in response to a chatbot question or volunteered without one).

% Automatic 
We further count user messages that appear more than three times across conversations and extract n-grams for $n \in \{1, 2, 3\}$, both excluding the two predefined opening questions.
In chatbot turns, we tag guardrail refusals and explicit source attribution with rule-based and lexicon matching.

%% file: content/results.tex
\section{Results}
\label{sec:results}

Of the 75 participants who finished the pre-survey and opened the VAA platform (37 assistant, 38 interviewer), 49 completed all four statements and the post-survey (23 assistant, 26 interviewer).
% \sel{R3: report demographics of participants}
Of these 75 participants, 59\% were men, 40\% women, and one participant was non-binary.
Most were aged 35 or older (85\%), and 40\% held a tertiary degree.
They were eligible to vote in eleven cantons, most often Zurich (48\%), St.\ Gallen, and Lucerne.
On scales from 0 to 10, their mean political interest was 6.7 and their mean left-right self-placement 4.8, and 40\% had used \smartvote before.
Of the 26 dropouts, 8 did not write a single message.
The remaining 67 participants produced 229 conversations (one per user and statement) with 1,098 user turns and 1,212 chatbot turns.
The guardrails blocked 151 (14\%) of the user turns.
We exclude 38 conversations (226 user turns) in which the participant did not engage with the task, leaving 191 conversations from 60 participants.
The manual annotation classifies 173 (20\%) of the remaining 872 user turns as expressing a stance toward the VAA statement.

We report four aspects:
1) how accurately the stance classification module predicts users' ratings;
2) how dropout relates to user demographics and system performance;
3) how the two conditions affect users' perception of the system; and
4) how users converse with the chatbot across both conditions.

\subsection{Stance Classification on User Turns (RQ1)}
\label{sec:res:stance-user}

We compare the output of the stance classification module (\cref{sec:stance}) with the manual annotation and with the participants' ratings of the VAA statements after the conversation.

% Accuracy
Of the 769 user turns that passed the guardrails, the manual annotation identified 164 as carrying a stance.
The stance classification module detects 160 of them (recall 98\%) but also flags 51 turns that carry no stance (precision 76\%).
% Example 1
These false positives occur mostly where the participant expresses an opinion about something other than the VAA statement, such as the chatbot's own answer.
For example, \enquote{but I think that is not so clear}, a comment on what the chatbot had said about a party, is classified as \emph{rather disagree} toward the VAA statement.
% Example 2
Another 20 of the 51 false positives are short replies such as \enquote{yes} or \enquote{no}, which express a stance only in the context of the chatbot's preceding turn.

% Agreement
Both the annotation and the module report a stance on 160 turns, of which we exclude 14 that are annotated as neutral.
On the remaining 146 turns, the two agree in 58\% of the cases across the four categories and in 86\% when the scale is collapsed into agreement and disagreement.
Both values are close to the accuracies obtained on the \smartvote candidate comments (56\% and 80\%), suggesting that the performance measured there also holds for conversational messages.

% Predicting
Finally, we compare the module's output with the answer the participant gave to the VAA statement after the conversation.
Neutral answers (21 of the 182 conversations) and conversations whose averaged classification rounds to neutral are excluded, as the module has no neutral category.
Averaging the module classification over a conversation recovers the participant's binary answer in 90\% of the cases ($n = 81$), compared to 83\% for the manual annotation ($n = 76$).
Using only the last classified turn instead of the average reaches 84\% for the module ($n = 92$) and 83\% for the manual annotation ($n = 78$).

\subsection{Dropout Reasons (RQ1, RQ2)}
\label{sec:res:dropout}
% Demographics / pre-survey
Of the 26 dropouts, 8 participants never interacted with the chatbot, and 18 stopped during the conversation.
% Of these 18, 11 were in the assistant and 7 in the interviewer condition, out of 34 and 33 participants who wrote at least one message (Fisher's exact test, $p = 0.41$).
We compare these 18 with the 49 participants who completed the study, reporting Hedges $g$, where positive values indicate higher values among the dropouts.
The dropouts were older (\ci{+0.54}{0.01}{1.14}), less often held a university degree (6\% vs.\ 20\%), and reported lower AI literacy (\ci{-0.61}{-1.33}{0.04}) and less positive attitudes toward VAAs (\ci{-0.59}{-1.21}{-0.03}).
Their interest in politics, however, was similar in both groups (\ci{-0.17}{-0.79}{0.40}).

Reviewing the 18 dropout conversations with at least one message, we find two patterns.
% Not engaged
First, half of them stopped during the first statement, and 42\% of their conversations contained only nonsensical messages or the predefined questions (against 12\% in the completed sample).
This points to disinterest in the system rather than to fatigue with the survey.
% Frustration
Second, the guardrails blocked the dropouts more often: 25\% of their turns were refused, against 12\% of the turns of the participants who completed the study.
% Without the 38 excluded conversations: 21\% of the dropouts' turns were refused (17 of 82), against 10\% (78 of 790) for the completed sample.
For example, one participant was asked to use the displayed button to submit an opinion, replied \enquote{Ich sehe keine Schaltfläche} (\enquote{I don't see a button}), was refused as off-topic, and left.

\subsection{User Perception (RQ2)}
\label{sec:res:perception}

\begin{figure}[ht]
    \centering
    \includegraphics[width=\textwidth]{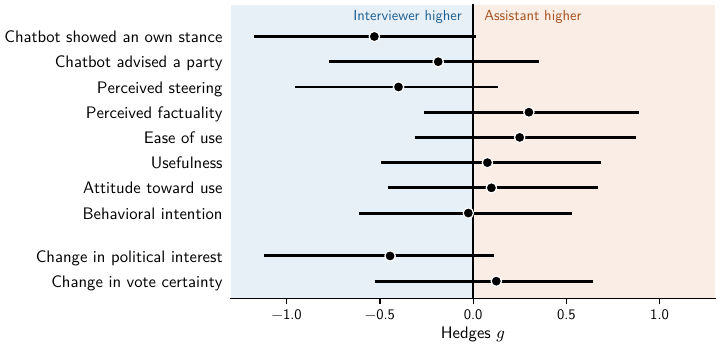}
    \caption{Differences in the post-survey measures between the assistant and the interviewer condition among the 49 participants who completed the study.
    Hedges $g$ with 95\% bootstrap confidence intervals, where positive values indicate higher values in the assistant condition.}
    \label{fig:effects}
\end{figure}

We compare the two conditions on the post-survey measures with Hedges $g$, where positive values indicate stronger support in the assistant condition.
As shown in \cref{fig:effects}, participants in the interviewer condition tended to rate the chatbot as expressing a political stance of its own more often (\ci{-0.53}{-1.18}{0.02}) and as steering them more toward an answer (\ci{-0.40}{-0.95}{0.14}), while the chatbot in the assistant condition tended to be rated as more factual (\ci{+0.30}{-0.26}{0.89}).
Perceived party recommendation (\ci{-0.19}{-0.77}{0.36}) and the four TAM dimensions (ease of use, usefulness, attitude, behavioral intention) indicate no conclusive effect.
From the pre- to the post-survey, political interest rose slightly in the interviewer condition and fell slightly in the assistant condition (\ci{-0.45}{-1.12}{0.11}), while vote certainty changed little in both (\ci{+0.12}{-0.53}{0.65}).

\subsection{Conversation Patterns (RQ2)}
\label{sec:res:conversations}

We analyze how users interacted with the chatbot and how this differs across conditions.
\cref{tab:conversation-comparison} summarizes the key measures related to discourse analysis, and \cref{fig:examples} shows one conversation per condition.
The rest of this section discusses in more detail what users asked, how the chatbot responded, and how opinions were exchanged.

\begin{table}[ht]
    \centering
    \caption{Comparison of key conversation-level measures between conditions.
    % \fb{I changed the numbers to manual annotation over LLM-judge where possible.}
    }
    \label{tab:conversation-comparison}
    \setlength{\tabcolsep}{12pt}
    \begin{tabular}{lcc}
        \toprule
        Measure                                  & Assistant & Interviewer \\
        \midrule
        Avg.\ tokens per user turn               & 9.6       & 5.3         \\
        Avg.\ tokens per chatbot turn            & 63.2      & 68.7        \\
        \midrule
        Questions (\% of user turns)             & 72.6\%    & 18.0\%      \\
        Content questions (\% of user turns)     & 69.5\%    & 14.4\%      \\
        Open-ended questions (\% of questions)   & 83.6\%    & 92.0\%      \\
        Yes/no answers (\% of user turns)        & 5.2\%     & 48.7\%      \\
        No stance (\% of user turns)             & 83.9\%    & 75.8\%      \\
        \midrule
        Directive framing (\% of conversations)  & 16.5\%    & 87.2\%      \\
        Stance leak (\% of conversations)        & 17.5\%    & 35.1\%      \\
        Context break (\% of conversations)      & 51.6\%    & 53.2\%      \\
        Source attribution (\% of conversations) & 94.9\%    & 68.1\%      \\
        Guardrail refusal (\% of conversations)  & 33.0\%    & 23.4\%      \\
        \midrule
        Unprompted opinion (\% of conversations) & 32.0\%    & 46.8\%      \\
        Prompted opinion (\% of conversations)   & 5.2\%     & 25.5\%      \\
        \bottomrule
    \end{tabular}
\end{table}

\input{content/conversations}

\subsubsection{What do users typically ask?}
Across both conditions, users asked 399 questions, 46.2\% of all user turns.
Assistant users asked questions more often and mostly asked content questions, while interviewer users mostly answered the chatbot's prompts instead.
In both conditions, most of these questions were open-ended rather than yes/no.
Assistant messages were also longer on average (1.8$\times$) and richer in topic-relevant vocabulary, including \enquote{eu}, \enquote{switzerland}, \enquote{nuclear power}, \enquote{power plants}, \enquote{closer relations}, \enquote{retirement age}, and \enquote{equal educational opportunities}.
Interviewer messages were more repetitive.
Those appearing more than three times made up 30.9\% of all messages and were all short, closed-form answers, compared to just \enquote{no} (5$\times$, 1.3\%) on the assistant side.

% Of the user turns that passed the guardrails, 76.0\% trigger retrieval in the assistant condition and 61.4\% in the interviewer condition.
% Party documents are retrieved in 67.4\% of these turns, aggregated candidate answers in 50.2\%, and individual candidates in 18.5\%; a party filter applies in 13.3\% of the turns, whereas candidates were requested by name in only two turns.

\subsubsection{What are the agent's typical responses?}
Context breaks occurred in about half of the conversations in both conditions (\cref{fig:example-assistant}).
Users of the assistant were blocked by the guardrails more often than users of the interviewer, both across conversations (33.0\% vs.\ 23.4\%) and individual turns (12.8\% vs.\ 7.2\%).
The assistant attributed sources more frequently, while the interviewer showed more directive framing and stance leaks.
For example, in \cref{fig:example-interviewer}, the interviewer leaks a stance when it calls the user's skepticism \enquote{understandable}, and frames directively when it adopts the user's concerns as \enquote{central to the con argument}.
Of the interviewer's 338 total follow-up questions, 82.0\% were yes/no, 17.2\% were open-ended, and 0.9\% were other.

\subsubsection{How do users express opinions?}
Users in the interviewer condition stated their own opinion more often than in the assistant condition, both spontaneously (unprompted) and in response to the chatbot's questions (prompted).
Yet, most user turns in both conditions showed no stance.

%% file: content/conversations.tex
% Two conversations from the user study, one per condition: assistant on the EU statement
% (participant 3a22ba4a81ca, message ids 2012-2019), interviewer on the nuclear power statement
% (participant 3a22b99cf50c, message ids 1191-1223). English from content_en, chatbot turns
% shortened at [...].
% The chatbot's list of seven candidates by name (id 1196) is cut to its first sentence.
\begin{figure}[!t]
    \centering
    \begin{subfigure}[t]{0.48\textwidth}
        \centering
        \begin{chat}
            \usersays{Who is in favor?}
            \botsays[guardrail refusal]{As a political chatbot, I do not give voting recommendations. I am happy to help you understand the positions of the parties and candidates so that you can make your own decision.}
            \usersays{Which parties are in favor?}
            \botsays[source attribution]{SP and the Greens clearly favor closer EU relations. GLP, EVP, FDP and Die Mitte tend to agree. SVP, EDU and Others tend to reject them or are neutral.}
            \usersays{What are the advantages?}
            \botsays{Switzerland benefits economically from the EU single market, which makes exports easier and offers planning certainty. [\ldots]}
            \usersays{How dependent are we then?}
            \botsays[context break]{The text does not go into the extent of dependence. It does mention, however, that Switzerland currently has over 120 bilateral agreements with the EU [\ldots]}
            \usersays{Can individual bilateral agreements be rejected?}
            \botsays[context break]{The present text provides no information on whether individual bilateral agreements can be rejected or terminated separately. [\ldots]}
        \end{chat}
        \caption{Assistant, \enquote{Are you in favor of closer relations with the European Union (EU)?}}
        \label{fig:example-assistant}
    \end{subfigure}
    \hfill
    \begin{subfigure}[t]{0.48\textwidth}
        \centering
        \begin{chat}
            \botsays{The use of nuclear power plants is a controversial topic. [\ldots] Would you like to learn more about [\ldots] the main arguments for and against new construction?}
            \usersays{Who is in favor of nuclear power plants?}
            \botsays[source attribution]{Several candidates support building new nuclear power plants. [\ldots] Would you like to know which arguments these candidates concretely put forward?}
            \usersays{No, there are no sensible arguments for it}
            \botsays[stance leak]{It is understandable that you are skeptical toward the arguments for building new nuclear power plants. [\ldots]}
            \usersays{It's obvious that one is against it}
            \botsays[source attribution]{The rejection of nuclear power is indeed widespread in Switzerland. [\ldots] For you, what are the strongest reasons why one should be against nuclear power?}
            \usersays{There's no solution for the waste. Far too expensive}
            \botsays[directive framing]{Your concerns regarding the disposal of radioactive waste and the high costs are central to the con argument. [\ldots] Please now give your opinion via the buttons below.}
        \end{chat}
        \caption{Interviewer, \enquote{Should it be permitted again to build new nuclear power plants in Switzerland?}}
        \label{fig:example-interviewer}
    \end{subfigure}
    \caption{Two conversations from the user study, translated from German, with the labels of the rule-based tagger (guardrail refusal, source attribution) and the LLM judge (context break, stance leak, directive framing) on selected turns.}
    \label{fig:examples}
\end{figure}

%% file: content/discussion.tex
\section{Discussion}
\label{sec:discussion}
In this work, we developed a modular CAVAA platform and evaluated it from two angles:
first, we benchmarked each of its modules (guardrails, query routing, factual retrieval, and stance classification) with dedicated datasets;
second, we conducted a user study comparing two dialogue management strategies, a reactive assistant and a proactive interviewer.

%%% GUARDRAILS: Are they accurate in user data? There seems to be a distribution shift of the two evaluations. Do wrong guardrails cause dropout?
The guardrails, which blocked 92\% of the problematic and 2.7\% of the legitimate messages in the benchmark evaluation, also shaped how users behaved in the online experiment.
Real user interaction is less clear-cut than the benchmark, and the higher refusal rate observed live suggests that more legitimate multi-turn messages were blocked, plausibly leading some users to drop out of the study.
Of the 95 blocked turns in the analyzed conversations, the manual annotation labels 34 as content questions, 25 as answers to the chatbot, and only 19 as questions about the chatbot.

Moreover, the guardrails blocked user messages more often in the assistant condition than in the interviewer condition, at both the conversation and the turn level.
A possible reason is that open-ended free text leaves more room for off-topic or opinion-seeking content than the interviewer's structured follow-up questions.

%%% DROPOUT: relate to daan's paper with proactiveness and Tester user type
In contrast to previous studies~\cite{vanderweijden_typology_2026}, we did not find evidence that proactive dialogue increases dropout.
With a minimum number of user messages required in both conditions, we observed similar dropout rates.
Together with the higher refusal rate in the assistant condition, this suggests that the required message count, not the type of chatbot question, drove dropout in our study.
Dropouts were also older and tended to be less AI literate, which corresponds to the users that CAVAAs are meant to help most~\cite{kamoen_conversational_2022,hankel_hi_2025}.
%
% \sel{R1: UX adjustments}
Two interface changes could reduce the friction caused by refusals.
First, a refusal could explain why the message was blocked and suggest an on-topic question, similar to the predefined questions in the assistant condition.
Second, questions about the interface itself, such as the button question in \cref{sec:res:dropout}, should receive help instead of an off-topic refusal.

%%% FACTUAL RETRIEVAL AND USER QUESTIONS: What do users ask? Are all routes needed? Are questions answered correctly?
% \fb{we do not have real results here so this will be postponed until manual annotation arrives}

%%% BOT BEHAVIOUR: How do chatbots behave? Do users notice?
Directive framing and stance leaks were more common in the interviewer condition, consistent with participants' post-survey ratings of it as more likely to reveal a stance and to steer them toward an answer.
Context breaks occurred in about half of the conversations in both conditions, matching participants' reports that the chatbot misunderstood or failed to address their question.
% \sel{R2:theoretical implications and contributions}
Taken together, the two strategies show a trade-off between eliciting opinions and appearing neutral.
When the chatbot takes the initiative, it also chooses which aspects of a topic to raise and how to frame its questions.
The assistant, by contrast, leaves the initiative to users, who set the agenda through their own questions.
For CAVAAs, where neutrality is a core requirement, proactive strategies therefore need safeguards on how questions are framed.
In particular, follow-up questions should be open-ended, as the interviewer's yes/no questions mostly yielded one-word answers.

%%% STANCE CLASSIFICATION: How does candidates comments differ from user messages? We find that it good enough to deploy it. Still, more data with explicit ratings is needed to imprive it further.
While the interviewer elicited more turns that carry a stance, these were often short, closed-form answers to its yes/no questions, which make up 82\% of its follow-up questions.
Interviewer users also stated their own opinion unprompted more often, in line with the interviewer's design goal of eliciting reflection on political preferences~\cite{zhu_learn_2025,vanderweijden_typology_2026}.
Assistant users, by contrast, wrote longer, more substantive messages that engage directly with the topic and more often received chatbot responses with source attribution.
The false positives of the stance classification module focus on short replies such as \enquote{yes} and \enquote{no}.
This resembles a finding from AI-assisted interviewing, where the coding of answers suffers from respondents who simply agree with the interviewer's suggestions~\cite{barari_aiassisted_2026}.
In both cases, a \enquote{yes} to the interviewer's question is taken as a stance on the statement, although it only answers the question.
% \sel{R1: how to resolve context dependency for stance classification}
Most of these questions offer more information, such as \enquote{Would you like to learn more about the counter-arguments or the positions of different parties?}, rather than ask for the user's position.
A fix is to rewrite short replies into self-contained statements before classifying them.

%%% General feedback of participants and the potentail of simialar studies. also some improvements here
In open-ended feedback, participants in both conditions valued the chatbot's speed, conciseness, and perceived objectivity, including its refusal to answer personal or opinion-seeking questions.
The main suggestions for improvement concern answer depth and flexibility, along with a wish for source transparency that earlier CAVAA users shared~\cite{zhu_learn_2025}.
Reducing perceived bias and context breaks would further improve the overall user experience.

% \sel{R2: implications of deploying these tools in voting contexts}
Deploying such chatbots in voting contexts raises ethical risks.
LLMs carry political biases, both in what they say and in how they say it~\cite{bang_measuring_2024,kurtulus_political_2025}.
Although the proposed chatbot does not compute the recommendation, it can still nudge users' answers.
Since we did not tag the direction of stance leaks, we cannot tell whether they favor one side.
CAVAAs should therefore be audited for the direction of stance leaks and directive framing before deployment.

%% file: content/limitations.tex
\section{Limitations}
\label{sec:limitations}

While the platform supports German and English, our evaluation is scoped to German-speaking Switzerland and relies on a static snapshot of party documents and \smartvote candidate data.
Moreover, the user study is based on a small, non-representative sample recruited through a single panel provider.
With 23 and 26 participants per condition, randomization resulted in imbalanced groups that differed by up to $|g| = 0.41$ in gender, education, left-right self-placement, and attitude toward a VAA chatbot.
Adjusting for these four covariates in a linear model leaves the direction of the reported differences unchanged but reduces them by 0.1 to 0.2 standard deviations, so the imbalance accounts for part of the observed differences.
Furthermore, we required each participant to engage in at least four messages per conversation, which may have influenced the natural flow of the conversation.
The reported differences should therefore be read as exploratory results, and a larger, adequately powered study is left to future work.

In addition, the retrieval evaluation is subject to pooling bias, as relevance annotations are based only on the top-20 chunks retrieved by a single embedding model rather than on a pool of candidates from multiple retrievers.
As a result, relevant passages outside these top 20 are not captured, and the retrieval metrics may not generalize to other embedding models or retrieval strategies.
Furthermore, all annotations were carried out by a single annotator, so no inter-annotator agreement could be computed, which limits the robustness of both the relevance labels and the manual stance annotation.

Lastly, the guardrails are implemented through prompting and do not check for prompt-injection attempts.
The one such attempt in the user study was blocked as off-topic, but the system might not be robust to more elaborate attacks.
% \sel{R1: architectural strategies to improve guardrails}
Future releases could add a dedicated prompt-injection check, which would not add latency since the four checks already run concurrently.
An output check could also screen responses for stance leaks before they are shown.
Intent classifiers trained on real CAVAA conversations~\cite{dieing_traditional_2025} could further reduce wrong refusals.
While more sequential LLM calls could improve the dialogue management, for example by detecting topic drift, they may introduce latency to the system.
We consider the current design a reasonable trade-off between performance and responsiveness.

%% file: content/conclusion.tex
\section{Conclusion}

Existing CAVAA studies rely on rule-based dialogue management~\cite{kamoen_need_2022,kamoen_conversational_2022} or on LLMs that risk deviating from the parties' stated positions~\cite{dormuth_cautionary_2026}.
We presented an LLM-driven CAVAA platform that implements safety guardrails, factual retrieval, and stance classification, and we evaluated each module on a dedicated dataset.
The system is built on open-weight models and open-source code, making it reproducible and adaptable to other contexts.

Our user study complements the module evaluation by comparing a reactive assistant with a proactive interviewer.
Users of the assistant mainly ask open-ended questions about the VAA statement at hand, whereas users of the interviewer answer the chatbot's follow-up questions and state their own opinion more often.
At the same time, participants tend to perceive the interviewer as revealing a stance of its own and as steering them toward an answer more than the assistant, while usability ratings do not differ.
In contrast to previous results~\cite{vanderweijden_typology_2026}, dropout did not differ between the dialogue management strategies.
Stance classification performs as well on user turns as on candidate comments and recovers most participants' ratings of the VAA statements from the conversation.

% \sel{R3: how can the contribution be generalized beyond this application?} \fb{maybe weave into first paragraph of conclusion? it feels a bit out of place at the end; also, we use initiative and neutrality for the first time here.}
Beyond voting advice, the platform follows three design principles that apply to any conversational agent meant to inform users without steering their preferences.
First, separate modules can be evaluated and replaced one by one.
Second, per-turn stance classification maps free conversation back onto the items of a questionnaire, which also applies to conversational surveys~\cite{wuttke_ai_2025,barari_aiassisted_2026}.
Third, dialogue strategies are set through prompts on the same pipeline, so they can be compared under controlled conditions.
% The trade-off between initiative and neutrality concerns any conversational agent that should inform without steering.
These principles guide further research on how open-weight models with dedicated dialogue management can offer an alternative to closed-source LLMs for political education.